\documentclass[12pt, fleqn]{iopart}
\usepackage{iopams}  

\begin{document}

\title[]{A fourth-order superintegrable system with a rational potential related to Painlev\'e VI}

\author{Ian Marquette$^{1}$, Sarah Post$^{2}$ and Lisa Ritter$^{2}$}
\address{$^1$School of Mathematics and Physics, The University of Queensland \\ Brisbane, QLD 4072, Australia}
\address{$^2$Department of Mathematics, University of Hawai`i M\={a}noa\\ Honolulu, HI 96815 USA}
\ead{lbritter@hawaii.edu}
\vspace{10pt}
\begin{indented}
\item[]\today
\end{indented}

\begin{abstract}
In this paper, we investigate in detail a superintegrable extension of the singular harmonic oscillator whose wave functions can be expressed in terms of exceptional Jacobi polynomials. We show that this Hamiltonian admits a fourth-order integral of motion and use the classification of such systems to show that the potential gives a rational solution associated with the sixth Painlev\'e equation. Additionally, we show that the integrals of the motion close to form a cubic algebra and describe briefly deformed oscillator representations of this algebra. 
\end{abstract}


\section{Introduction}
There has recently been much interest in exceptional orthogonal polynomials \cite{gomez2009extended, GUKM2010}. These are families of orthogonal polynomials with gaps in their degree. If we consider extensions of the classical orthogonal polynomials, these gaps in degree allow for a way to extend B\"ochner's theorem beyond the classical case. In particular, these exception orthogonal polynomials will satisfy a Sturm-Liouville problem
\begin{equation}
	A(x) f''(x) + B(x) f'(x) + C(x) f(x) + \lambda_{n} f(x) = 0,
\end{equation}	
where coefficient functions are allowed to be rational. An important avenue for this current work is in the connection between these polynomials and Painlev\'{e} equations. The connection between classical OPs and Painlev\'e equations has been well studied \cite{KKY1999, vanassche2017} and 
recent work has been done to connect these exceptional polynomials to transcendents of the Painlev\'{e} IV and V differential equations \cite{CGGM2018}.

One surprising area where these connections arise is in the study of superintegrable systems. For example, the connection between the exceptional Hermite equations and Painlev\'e IV is accomplished by considering rational extensions of the harmonic oscillator, the canonical example of an exactly-solvable system\cite{MQ2015}. In higher dimensions, the harmonic oscillator is an example of a superintegrable system i.e. a Hamiltonian system with  $2d-1$ integrals of motion \cite{MWP2013}.  Superintegrable systems are closely associated with classical orthogonal polynomials and extensions of these systems have been constructed by swapping out the classical eigenvalue operator for the exceptional one \cite{post2012families, MQ2013}. If the classical superintegrable system was multiseparable, this extension is usually separable in only one orthogonal coordinate system but remains superintegrable with an additional integral of higher-order. 

Research into superintegrable systems that are separable and admit a higher-order symmetry (more than quadratic in the momenta) began to be systematically investigated by Winternitz and his students in the early 2000's \cite{Gravel2003, tremblay2010third}. Remarkably, Painlev\'e transcendents in their full generality arose in systems that separate in Cartesian or polar coordinates with an additional integral of motion that is third-order in the momenta.  More recently this analysis has been extended to separable systems with a fourth-order integral. Painlev\'e VI has also been associated with fourth-order superintegrable systems that separate in polar coordinates \cite{escobar2017fourth}.

 In this work, we consider a single instance of the family of Hamiltonians defined in \cite{post2012families}. This Hamiltonian admits separation of variables in polar coordinates with the angular function comprised of an exceptional Jacobi polynomial. We will show that there is an additional fourth-order integral and so the system falls under the classification of Escobar et al \cite{escobar2017fourth}. Indeed, this system is a special case of a more general potential related to the sixth Painlev\'e equation. Finally, we present the algebra generated by the integrals of motion which turns out to be a cubic algebra. 
			
\section{The Hamiltonian}
We consider the following Hamiltonian from \cite{post2012families},
\begin{equation}H = - \Delta+ \omega^2 r^{2} + \frac{1}{ r^{2}}  \left( \frac{ \alpha^{2} - \frac{1}{4}}{\cos^{2}{\phi}} + \frac{\beta^{2} -\frac{1}{4} }{\sin^{2}{\phi}} + \frac{8(1+ b \cos{(2 \phi}))}{(b+ \cos{2 \phi})^{2}} \right)\label{H}
\end{equation}
with $b = \frac{\alpha + \beta}{\beta-\alpha}$.  This potential is a deformed Smordinsky-Winternitz potential \cite{SW1965} with a rational deformation that scales with $\hbar^2.$ While the original paper considers a family of deformation depending on a rational k, we focus only on the $k=1$ case. Other choices of rational $k$ will lead to higher-order symmetry operators and are not necessarily of interest for the current study because the choice of $k$ does not affect the rational extension in the angular variable. 

We begin with a brief description of the separated wave functions for the associated Schr{\"o}dinger equation
\begin{equation}
	H \Psi - E \Psi = 0.
\end{equation}
This equation separates in polar coordinates as 
\begin{equation}
	\Psi = \Phi(\phi) R (r)
\end{equation}
with 
\begin{equation}
	\label{radial} \left( - \frac{1}{r} \partial_{r} r \partial_{r} + \omega r^{2} + \frac{ C^{2}}{r^{2}} \right) R(r) -ER(r)=0
\end{equation}

\begin{equation}\label{angular}
	\left( - \partial_{\phi}^{2} + \frac{ \alpha^{2} - \frac{1}{4}}{ \cos^2{\phi}} + \frac{ \beta^{2} - \frac{1}{4}}{ \sin^{2}\phi} + \frac{8 ( 1 + b \cos{2 \phi})}{ ( b+ \cos{2 \phi} )^{2}} - C^{2} \right) \Phi(\phi) =0.
\end{equation}
The radial equation (\ref{radial}) is solved by the following change of variable and conjugation by the ground state
\begin{equation}
y=\omega r^2, \qquad 
R_0(y)= y^{C/2}e^{-y/2}. \end{equation}
Resulting in the following eigenvalue equation for Laguerre polynomials
\begin{equation}
\left[y\frac{d^2}{dy^2} +(C+1-y)\frac{d}{dy}+m\right] F(y)=0,
\end{equation}
where $m$ is associated with energy quantization 
\[ E= \omega(2m+C+1).\] 
The radial part of the eigenfuctions then become
\[ R_m^{C}(y)=  y^{C/2}e^{-y/2} L_m^{C}(y),\qquad  m=0,1,2, \ldots.\]

For the angular equation, we make use of the fact that it factorizes as 
\[  - \partial_{\phi}^{2} + \frac{ \alpha^{2} - \frac{1}{4}}{ \cos^2{\phi}} + \frac{ \beta^{2} - \frac{1}{4}}{ \sin^{2}\phi} + \frac{8 ( 1 + b \cos{2 \phi})}{ ( b+ \cos{2 \phi} )^{2}} =AA^\dagger+(\alpha-\beta-1)^2,\]
where 
\[ A= \partial_{\phi} - \frac{\xi'(\phi)}{\xi(\phi)}, \qquad \xi(\phi) =(\cos \phi)^{-\alpha-\frac12}(\sin \phi)^{\beta-\frac12}(b+\cos (2\phi)).\]
Note that the last factor of the seed function $\xi$ is proportional to a degree 1 Jacobi polynomial $P_1^{-\alpha-1, \beta-1}(-\cos(2\phi)).$
The reverse order of the factorization gives the operator
\[ A^\dagger A+(\alpha-\beta-1)^2= - \partial_{\phi}^{2} + \frac{ (\alpha+1)^{2} - \frac{1}{4}}{ \cos^2{\phi}} + \frac{ (\beta-1)^{2} - \frac{1}{4}}{ \sin^{2}\phi},\]
whose eigenfunctions can be expressed as 
\begin{equation}
\label{L2neeign} \Psi_n(\phi)=(\cos\phi)^{\alpha+\frac32}(\sin \phi)^{\beta-\frac12}P_n^{\alpha+1, \beta-1}(-\cos(2\phi)),
\end{equation}
with eigenvalues 
\begin{equation} \label{cn} C_{n}^{2}=(2n+\alpha+\beta+1)^2. \end{equation}
This makes the energy of the full Hamiltonian is 
\begin{equation}
E = \omega (2 + 2m + 2n + \alpha +\beta) \label{energy}
\end{equation}
Making use of the factorized form, we can then express the wavefunctions for the original operator (\ref{angular}) as 
\begin{equation}\label{Phi} \Phi_n(\phi)=A \Psi_n. \end{equation}

To explicitly connect these wave functions to the exceptional orthogonal polynomials, we make the following change of variable,
\begin{equation}
	x = -\cos{(2 \phi)}
\end{equation}
and conjugate by a gauge factor
\begin{equation}
	G_{x} = \frac{(1-x)^{\frac{\alpha}{2} + \frac{1}{4}} (1+x)^{\frac{\beta}{2}+\frac{1}{4}}}{(x-b)}.\label{Gx}
\end{equation}
To insure our gauge factor has the proper form we note that both $\alpha$ and $\beta$ must be greater than $-\frac{1}{2}$.
With this change of variable and gauge factor, the angular function can be expressed as 
\[ \Phi_n(\phi) = X_{n}(x)G_x, \quad n\ge 1.\]
The function $X_n$ will be eigenfunctions of the following operator
\begin{equation}
	T^{\alpha, \beta} = 4(x^{2} -1) \partial_{x}^{2} + \frac{4 (\beta - \alpha) ( 1-b x)}{b-x} \left( ( x - b+\frac{2}{\alpha-\beta}) \partial_{x}-1 \right) + ( \alpha + \beta + 1)^{2}.
\end{equation}
Comparing this equation to the one given in \cite{gomez2009extended}, we see that the $X_n$ are exceptional Jacobi polynomials given by 
\[ X_n=\hat{P}_n^{\alpha, \beta}(x).\]
Note that there is no constant eigenfunction and so the gauge factor (\ref{Gx}) is not the ground-state. Instead the lowest energy state ($n=0$) contains a degree 1 polynomial. 

\section{Integrals of Motion}
In this section we present the integrals of motion and describe the nonlinear equation satisfied by this potential. 

By construction,  we see that the system is separable in polar coordinates and so it admits the following second-order integral: 
\begin{equation}
L_1= - \partial_{\phi}^{2} + \frac{ \alpha^{2} - \frac{1}{4}}{ \cos^2{\phi}} + \frac{ \beta^{2} - \frac{1}{4}}{ \sin^{2}\phi} + \frac{8 ( 1 + b \cos{2 \phi})}{ ( b +\cos{2 \phi} )^{2}} .
\end{equation}

We then used computer software (MAPLE and Mathematica) to search for a higher-order integral for this system and compared our results with the classifications in \cite{tremblay2010third, escobar2017fourth}. In agreement with the results in \cite{tremblay2010third}, there are no third-order integrals for this confining potential. On the other hand, there is a fourth-order integral in agreement with \cite{escobar2017fourth}. 
Note that our potential is of the form 
\[ V=\omega^2 r^2+\frac{2}{r^2}T'(\phi)\]
and so according to the classification, will have an integral of motion of the form
\begin{equation} L_2=\{p_x^2-p_y^2, L_z^2\}+\{g_1, p_x^2\}+\{g_2, p_xp_y\} +\{g_3, p_y^2\} +G_0.
\end{equation}
It is useful to change the form of the integral to polar coordinates and collect the undetermined functions as 
\[ L_2=\{p_x^2-p_y^2, L_z^2\}-2  \left( G_1 \partial_r^2+G_2\partial_r\partial _\phi+G_3\partial_\phi^2\right)+ \ldots\]
The determining equations are then  derived from the odd degree coefficients of the equation 
\[ [H, L_2]=0.\]
For a general description of the procedure see \cite{escobar2017fourth}. 

The first set of differential equations involve only the $G_i$'s  and are
\begin{eqnarray}
\partial_r G_1=0\\
r^2\partial_r G_2 +\partial_{\phi} G_1=2\cos(2\phi)T''\\
r^3\partial_r G_3 +r \partial_\phi G_2+2G_1=-6\sin(2\phi) T''-4\cos(2\phi)T' +4\cos(2\phi)\omega^2r^4\\
                    r^{2} \partial_\phi G_3+rG_2=-4\cos(2\phi) T''+4\sin(2\phi) T' -4\sin(2\phi)\omega^2r^3.
\end{eqnarray}
Solutions of these equations are given by 
\begin{eqnarray*}
 G_1= 2\cos(2\phi)T'-2\sin(2\phi)T -\frac{c_{12}}{4}\cos(2\phi)+\frac{c_{11}}{4}\sin(2\phi)+c_{10}\\
G_2=\frac{-1}{2r}\left( 12\sin(2\phi)T'+8\cos(2\phi)T-c_{12} \sin(2\phi) -c_{11}\cos(2\phi) \right)\\ +\frac{c_{22}}2\cos(\phi)-\frac{c_{11}}2\sin(\phi)
\\
G_{3}=\frac{-1}{4r^2}\left(16\cos(2\phi)T' -8\sin(2\phi)T -c_{12}\cos(2\phi)+c_{11} \sin(2\phi)-2c_{10}\right)\\ \qquad +\frac{-1}{2r} \left(c_{21}\cos(\phi) +c_{22}\sin(\phi) \right) +2\cos(2\phi)\omega^2r^2+\frac{c_{30}}{2}
\end{eqnarray*}
with $c_{10}, c_{11}, c_{12}, c_{22}, c_{21} $ and $c_{33}$ arbitrary constants. 
The remaining equations to solve are
\begin{eqnarray}
\label{G0r}
\fl r^3\partial_rG_0= \frac{\sin(2\phi)}{2}T^{(4)}+4\cos(2\phi)T^{(3)}
\\ \fl +\left(-3\sin(3\phi)T'-2\cos(2\phi)T-10\sin(2\phi) +\frac{c_{11}}{4}\cos(2\phi)+\frac{c_{12}}{4} \sin(2\phi)+\frac{c_{21}r}{4} \sin(\phi)+\frac{c_{22}r}{4}\cos(\phi)\right)T''\nonumber\\ \fl 
+\left(-4\cos(2\phi) T'+4\sin(2\phi)T+4\omega^2r^4\cos(2\phi)+\frac{c_{12}}{2}\cos(2\phi)-\frac{c_{11}}{2} \sin(2\phi)-8\cos(2\phi)-c_{10}\right)T' \nonumber
\\
\fl -\frac{\omega^2r^4}{2}\left(8\sin(2\phi)T-c_{11}\sin(2\phi)+(c_{12}+16)\cos(2\phi) -2c_{10}\right)\nonumber
\end{eqnarray}
and 
\begin{eqnarray}
\label{G0t}
\fl r^2\partial_{\phi}G_{0} = -\sin(2\phi) T^{(3)}\\
\fl +\left( \cos(2\phi)(\frac{c_{12}}{4} +2\omega^2r^4-4T')+\sin(2\phi)(2T-\frac{c_{11}}{4})  +\frac{c_{21}}{2} r\cos(\phi) -\frac{c_{22}}{2} r\sin(\phi) +\frac{c_{33}}{2}r^2+\frac{c_{10}}{2} \right) T''\nonumber\\
\fl +\left( 6 \sin(2\phi) T' +4\cos(2\phi) T -\frac{c_{11}}{2} \cos(2\phi) -(\frac{c_{12}}2 +6\omega^2r^4)\sin(2\phi) -\frac{c_{21}}2 r\sin(\phi) -\frac{c_{22}}{2} r\cos(\phi)\right)T' \nonumber\\
\fl + \frac{\omega^2 r^4}{2} \left( (c_{11}-8T) \cos(2\phi) +(c_{12}+16) \sin(2\phi) +c_{21}r \sin(\phi) +c_{22} r\cos(\phi) \right)\nonumber
\end{eqnarray}

The compatibility equations of this separate in powers of $r$. One of which gives $c_{12}=c_{22}=0$ and the other gives a nonlinear equation for $T$: 
\begin{eqnarray*} & 12\tan(2\phi)\left(T''\right)^2+\tan(2\phi) T^{(5)} +10T^{(4)}\\
& +(-12\tan(2\phi)T'-8T+\tan(2\phi)(c_{12}-40)+c_{11})T^{(3)}\\
&+(48\tan(2\phi)T-96T'-6\tan(2\phi)c_{11}+6c_{12}-80)T'' +96\tan(2\phi)(T')^2\\
&+(64T+64\tan(2\phi)-8c_{11}-8\tan(2\phi)c_{12})T'=0.\end{eqnarray*}
The general solution of this nonlinear differential equation was found in \cite{escobar2017fourth}. For our particular choice of a potential, 
\[T=\frac14(\alpha^2-\frac14)\tan(\phi)-\frac14(\beta^2-\frac14)\cot(\phi)+\frac{4 \sin(2\phi)}{b+\cos(2\phi)},\] the constants are 
\[ c_{11}=0, \qquad c_{12}=8(\alpha^2-\alpha\beta+\beta^2)+18.\]
With these constants, the equations (\ref{G0r}-\ref{G0t}) are consistent and can be solved to finally determine the fourth-order integral. The function is contained in the Appendix. The remaining constants $c_{10}$ and $c_{30}$ are related to the freedom to add multiplies of the Hamiltonian and second-order integral to the fourth-order integral. 

\section{The nonlinear equation}
Following the derivation in Ref. \cite{escobar2017fourth}, we express the function $T$ as
\[ T(\phi)=\frac{-2 W(\phi)}{\sin(\phi)\cos(\phi)} -\frac{\alpha^2-\alpha\beta +\beta^2+\frac{7}{4}}{\tan(2\phi)},\]
and making the change of variables $y= \frac12 (1+\cos(2\phi))$ gives the following nonlinear equation for $W(y):$
\[ y^2(1-y)^2(W")^2+4W'(yW'-W)^2-4(W')^2(yW'-W)
\]
\begin{equation}\label{SD1}
+4q_7(W')^2 +4q_8W'+4q_9(yW'-W)+4q_{10}=0.\end{equation}
This is  the equation SD-I.a in the classification by Cosgrove \cite{highpainCosgrove} with constants 
\[ q_1=q_4=q_5=q_6=0,\qquad  q_2=-q_3=1, \qquad 4q_7=-(\alpha^2-\alpha\beta+\beta^2+2)\]
and 
\[ 4q_8=\frac{1}{4}(\alpha+\beta)(\alpha -\beta)(\alpha-\beta+2)(\alpha-\beta-2)\]
\[4q_9=\frac14(2\alpha^2+\beta^2)(\alpha-\beta)^2-\alpha\beta-\beta^2-1\]
\[4q_{10}=-\frac1{16}(\alpha^2+\alpha\beta+\beta^2-2)(\alpha-\beta)^2-\frac12(\alpha^2+\beta^2).\]

The solution of this equation can be expressed in terms of Painlev\'e VI with relation between the parameters
\[-4q_7 =- \gamma_1- \gamma_2+ \gamma_3-  \gamma_4 - \sqrt{2\gamma_1} + 1\]
\[ -4q_8 = (\gamma_2 + \gamma_3)(\gamma_1 + \gamma_4 - \sqrt{2\gamma_1}),\]
\[ -4q_9 = (\gamma_3 - \gamma_2)(\gamma_1 - \gamma_4 - \sqrt{2\gamma_1} + 1) + \frac14
(\gamma_1 - \gamma_2 - \gamma_3 + \gamma_4 -\sqrt{2\gamma_1})
^2\]
\[ -4q_{10} =\frac14
(\gamma_3 - \gamma_2)(\gamma_1 + \gamma_4 - \sqrt{2 \gamma_1})^2+\frac14(\gamma_2 + \gamma_3)^2(\gamma_1 - \gamma_4- \sqrt{2\gamma_1} + 1).\]
A solution for this system is given by 
\begin{eqnarray}
\gamma_1=\frac12 (\alpha-\beta \pm 1)^2,\qquad 
\gamma_2= -\frac12 \alpha^2, \qquad \gamma_3=\frac12 \beta^2,  \qquad \gamma_4=\frac{-3}{2}.
\end{eqnarray}
In terms of the seed function, the function $W$ is given by
\[\fl W=\frac{(\alpha-\beta)^2}{8}(1-2y)+\frac{(\alpha+\beta)(\alpha-\beta+2)(\alpha-\beta-2)}{8(\alpha-\beta)}+\frac{1}{(\alpha-\beta)^2}\partial_y \log P_1^{-\alpha-1, \beta-1}(1-2y) 
.\]

\section{The Algebra}

The system admits a cubic algebra with generators $H$ as in (\ref{H}) and
\[ X=\frac{1}{2}(L_1-1), \qquad Y=\frac{1}{8}L_2,\]
as derived in the previous section. See the Appendix for the explicit coefficient functions. We make use of the freedom in $Y$ to add constant multiples of $H$, $X$ and the identity to simplify the algebra. In particular, we take $c_{33}=c_{10}=0.$

The cubic algebra takes the following simple form, see \cite{marquette2015deformed}
\begin{eqnarray*}
	\left[ X,Y \right] = Z\\
	\left[ X,Z \right] =  b \{X , Y\}  + f \\
	\left[ Y,Z\right] = - b Y^{2} +gX^{3} + h X^{2}  + iX+j
\end{eqnarray*}
with
\begin{eqnarray*}
b=16, \qquad  f= 2\, (\alpha+\beta)(\alpha-\beta) (\alpha-\beta+2)(\alpha-\beta-2)H\\
g=-1024\, \omega^2, \qquad h=48 H^2+256\omega^2(3\alpha^2-3\alpha\beta+3\beta^2-4)\\
i=-32(\alpha^2-\alpha\beta+\beta^2)H^2-192\omega^2(\alpha^2+\beta^2-4)(\alpha-\beta)^2\\
j=\left(6(\alpha^2+\beta^2-4)(\alpha-\beta)^2+16(\alpha^2-\alpha\beta+\beta^2)\right)H^2\\\qquad +16\, \omega^2(\alpha-\beta+2)(\alpha-\beta-2)(\alpha-\beta)^2(\alpha^2+\alpha\beta+\beta^2-4).
\end{eqnarray*}

It is interesting to note that these constants are linear if written in terms of the parameters of Equation (\ref{SD1}). 
\begin{eqnarray*}
b=16, \qquad f=32q_8H\\
g=-1024\omega^2, h=48H^2-512\, \omega^2(6q_7+5)\\
i=64(2q_7+1)H^2-512\omega^2(3+8q_7-2q_8-4q_9)\\
j=16(1+4q_7-2q_8-4q_9)H^2-256\omega^2(1+4q_7-2q_8-4q_9+4q_{10}).
\end{eqnarray*}
The Casimir operator has the form
\begin{eqnarray*}
	K = Z^2 - b \{X,Y^{2}\} +b^{2} Y^{2} - f Y \\ \qquad + \frac{g}{2} X^{4} + \frac{2}{3} ( h + g b)X^{3} + \left( -\frac{1}{6} g b^{2} + \frac{b h}{3} + i\right) X^{2} 
	+ 2j X.
\end{eqnarray*}

With the operator put in, this simplifies to 
\begin{equation}
	K = K_2H^{2} +K_0
\end{equation}
with
\begin{eqnarray}
	K_2 = 64 \left(  \left( {\alpha}^{2}+\alpha\,\beta+{
\beta}^{2}-2 \right)  \left( \alpha-\beta \right) ^{4}+8\,{\alpha}^{2
}+8\,{\beta}^{2} \right) 
  \\
K_0=2{\omega}^{2} \left(\alpha -\beta-2 \right) 
^{2} \left(\alpha -\beta+2 \right) ^{2} \left( \alpha+\beta \right) ^
{2} \left(\alpha -\beta \right) ^{2}.
\end{eqnarray}

\section{Spectrum Calculations}
Some results on realization and representations of polynomial algebras with three generators are known. Here we consider the approach relying on deformed oscillator algebras \cite{MQLissajous} \cite{Das2001}. In this approach, we assume that one operator is diagonalized and acts as a constant depending on some quantum number $n$. The other operator is assume to be at most three-terms, shifting the quantum number up or down by 1. 

We construct the operators making use of an oscillator algebra generated by the number operator $N$, which acts as multiplication by $n$ in the representation, and generators $b,$ $b^{\dagger}$ which shift the number up and down by 1, respectively. These relations are encode in the following algebra relations:
\[ [N,b]=-b,\quad  [N,b^{\dagger}]=b^{\dagger} \]
\[ b^{\dagger}b=\Phi(N),\quad bb^{\dagger}=\Phi(N+1) \]
where $\Phi(N)$ is the structure function. 

The generators of our algebra are then expressed as 
\[ X=X(N), \qquad Y= Y_0(N) +b^\dagger \rho(N)+\rho(N)b.\]
Where the functions of $N$ can be explicitly defined in our case:
\[ X(N) = 4(N+u)^{2} -1\]
\[ Y_{0}(N) = \frac{H(2-\alpha+\beta)(\alpha-\beta)(2+\alpha-\beta)(\alpha+\beta)}{(N+u)^{2}-\frac{1}{4}}\]
\[ \rho (N) = \frac{1}{2048 \sqrt{6(N+u)(N+u+1)(1+2(N+u))^{2} \beta^{9}}}\]

The cubic algebra relations give the structure function as 
%
\[ \Phi(u,H,N)=\]
\[3\cdot 2^{39}(\alpha-\beta-2 N-2u-1) (\alpha-\beta-2 N-2u+1) (\alpha-\beta-2 N-2u+3) \]
\[ (\alpha-\beta+2 N+2u-3) (\alpha-\beta+2 N+2u-1) (\alpha-\beta+2 N+2u+1) (\alpha+\beta-2 N-2u+1)\]
\[ (\alpha+\beta+2 N+2u-1) (H^2-2 \hbar^2 (1-2 N-2u)^2 \omega^2) \]
where $u$ is a representation dependent constant. In terms of the energy, the expression can be re-written as 
\[ \Phi(u,E,n)= -3\cdot 2^{39}  \omega^2 \]
\[(n+u-\frac{1}{2} (1-\alpha-\beta))
(n+u-\frac{1}{2} (-1+\alpha-\beta))\]
\[(n+u-\frac{1}{2} (3+\alpha-\beta))
(n+u-\frac{1}{2} (1+\alpha-\beta))\]
\[(n+u-\frac{1}{2} (-1-\alpha+\beta))
(n-\frac{1}{2} (1-\alpha+\beta))\]
\[(n+u-\frac{1}{2} (1+\alpha+\beta))
(n+u-\frac{1}{2} (3-\alpha+\beta))\]
\[(n+u-\frac{(-\sqrt{2}E+2\hbar \omega)}{4 \hbar \omega})
(n+u-\frac{(\sqrt{2}E+2\hbar \omega)}{4 \hbar \omega}).\]
The existence of finite dimensional unitary representation provide the following constraints on the structure function $\Phi(u,E,n)$ where $n+u$ are eigenvalues of $N$ and $E$ is the spectrum: 
\[  \Phi(u,E,0)=0, \quad \Phi(u,E,p+1)=0 \]
and positive for $n=1,...,p$ . The first constraints provide the following solutions
\[ u_{1}=\frac{1}{2}( \epsilon_{1} + \epsilon_{2}(\alpha-\beta) ) \]
\[ u_{2}=\frac{1}{2}( 3 + \epsilon_{1}(\alpha_{1}-\beta_{1} ) ) \]
\[ u_{3}=\frac{1}{2}( 1+ \epsilon_{1} (\alpha+\beta )), \]
where $\epsilon_{1,2}=\pm 1.$ 
The second constraints corresponds to a quantized energy spectrum that, given the choices of $u_{i}$'s above, reduce to 
\[E_{1} = \omega ( 1 + 2 p  + \epsilon_{1} + \alpha \epsilon_{2} - \beta \epsilon_{2} )   \]

\[ E_{2} = \omega ( 4 + 2 p  +  \alpha \epsilon_{1} - \beta \epsilon_{1} )\]
		
\[ E_{3} = \omega ( 2 + 2 p  +  \alpha \epsilon_{1} + \beta \epsilon_{1} ) .    \]		
These energy levels match what is seen in Equation \ref{energy}, where iterative over $p$ takes the place of $n$ and $m$.  The epsilons also pick up the energy levels of the Hamiltonians that have different choices for the signs of $\alpha$ and $\beta$.

\section{Conclusion}
In this paper, we closely examine the $k=1$ Hamiltonian given  in \cite{post2012families}. We affirm that this Hamiltonian is superintegrable and show that it has a integral of motion of order 4, in addition to the integral of motion of order 2 associated with separation of variables. Thus, this Hamiltonian belongs in the classification of \cite{escobar2017fourth} and the rational potential satisfies a non-linear differential equation which can be transformed into Painlev\'e VI. In addition, we show that the integrals form a cubic algebra, which can be modeled using a deformed algebra approach. We give the structure functions for such representations. 

In a future paper, we will discuss rational deformations of the potential corresponding to the full family of exceptional polynomials. We anticipate that the members of the full family of exceptional orthogonal polynomials will have a similar structure and will give rational solutions of the non-linear equation of arbitrary order. The precise connection between the rational solutions for (\ref{SD1}) and Painlev\'e VI will be further investigated with the goal of presenting closed, rational solutions of Painlev\'e VI.


\section*{Bibliography}
\bibliographystyle{unsrt}
\bibliography{exceptional}

\appendix
\section{The constant coefficient}
The final function in the fourth-order operator is as follows:
\begin{eqnarray} \fl G_0=2\omega^2 r^2\bigg(\frac{\frac14-\alpha^2}{y} +\frac{\frac14-\beta^2}{y-1} -\frac{8\alpha\beta(\alpha+\beta)}{((\alpha-\beta) y-\alpha)^2(\alpha-\beta)}-\frac{4(\alpha^2+4\alpha\beta+\beta^2)}{((\alpha-\beta) y-\alpha)(\alpha-\beta)}\nonumber \\ 
-2((\alpha-\beta)^2+3)y +2\alpha(\alpha-\beta)-\frac{\alpha-7\beta}{\alpha-\beta}\bigg)\nonumber\\
\fl +r^{-2}\bigg(\frac{(1-4\alpha^2)(-8\alpha^3+8\alpha^2\beta+12\alpha-32\beta)(y)+\alpha(1-4\alpha^2)}{8\alpha y^2}\nonumber\\
+\frac{(1-4\beta^2)(-8\beta^3+8\alpha\beta^2+12\beta-32\alpha)(y-1)+\beta(1-4\beta^2)}{8\beta (y-1)^2}\nonumber\\
+\frac{1}{{\alpha\beta (\alpha-\beta) y-\alpha)^4}}\bigg(-{
\alpha}^{3} \left( 4\,{\alpha}^{2}{\beta}^{2}-{\alpha}^{2}-3\,\alpha\,
\beta+5\,{\beta}^{2} \right)\nonumber\\
+{\alpha}^{2} \left( 12\,{
\alpha}^{3}{\beta}^{2}-12\,{\alpha}^{2}{\beta}^{3}-3\,{\alpha}^{3}-12
\,{\alpha}^{2}\beta+8\,\alpha\,{\beta}^{2}+23\,{\beta}^{3} \right) y\nonumber\\
-\alpha\, \left( 12\,{\alpha}^{4}{\beta}^{2}
-24\,{\alpha}^{3}{\beta}^{3}+12\,{\alpha}^{2}{\beta}^{4}-3\,{\alpha}^{
4}-15\,{\alpha}^{3}\beta-22\,{\alpha}^{2}{\beta}^{2}+53\,\alpha\,{
\beta}^{3}+3\,{\beta}^{4} \right) {y}^{2}\nonumber\\
 \left( \alpha-\beta \right)  \left( 4\,{\alpha}^{4}{\beta}^{2}-8\,{
\alpha}^{3}{\beta}^{3}+4\,{\alpha}^{2}{\beta}^{4}-{\alpha}^{4}-7\,{
\alpha}^{3}\beta-32\,{\alpha}^{2}{\beta}^{2}-7\,\alpha\,{\beta}^{3}-{
\beta}^{4} \right) {y}^{3}\bigg)\nonumber
\end{eqnarray}

\section{Structure functions}
The following are the structure functions for the 3 different choices of parameter $u_i$: 
\[ \Phi_{1}=3\cdot 2^{39}  \omega^2 \omega^2  (p+1-N) (N+p  + \epsilon_{1}+ \alpha \epsilon_2 - \beta \epsilon_2 ) \]

\[\frac{1}{2} (\alpha \epsilon_2+\alpha-\beta \epsilon_2+\beta_1+2 N+\epsilon_1-1) \frac{1}{2} (\alpha (\epsilon_2-1)-\beta \epsilon_2+ \beta+2 N+\epsilon_1+1)  \]

\[ \frac{1}{2} (\alpha (\epsilon_2-1)-\beta \epsilon_2+\beta+2 N+\epsilon_1-3) \frac{1}{2} (\alpha (\epsilon_2-1)-\beta \epsilon_2+\beta+2 N+\epsilon_1-1) \]

\[ \frac{1}{2} (\alpha (\epsilon_2+1)-\beta (\epsilon_2+1)+2 N+\epsilon_1+1) \frac{1}{2} (\alpha ( \epsilon_2+1)-\beta (\epsilon_2+1)+2 N+\epsilon_1-1) \]

\[ \frac{1}{2} (\alpha (\epsilon_2-1)-\beta (\epsilon_2+1)+2 N+\epsilon_1-1)\frac{1}{2} (\alpha (\epsilon_2+1)-\beta (\epsilon_2+1)+2 N+\epsilon_1)-3)\]

\[ \Phi_{2}= 3\cdot 2^{39} \omega^2  (p+1-N) (\alpha  \epsilon_1-\beta  \epsilon_1+N+p+3) \]

\[ \frac{1}{2} (\alpha \epsilon_1+\alpha -\beta \epsilon_1+\beta +2 N+2) \frac{1}{2} (\alpha  (\epsilon_1-1)-\beta \epsilon_1+\beta+2 N+4) \]

\[ \frac{1}{2} (\epsilon_1-1) (\alpha -\beta)+N)  \frac{1}{2} (\alpha  (\epsilon_1-1)-\beta \epsilon_1+\beta+2 N+2)  \]

\[ (\frac{1}{2} (\alpha  (\epsilon_1+1)-\beta  (\epsilon_1+1)+4)+N) (\frac{1}{2} (\alpha  (\epsilon_1+1)-\beta (\epsilon_1+1)+2)+N)\]

\[ (\frac{1}{2} (\alpha  (\epsilon_1-1)-\beta  (\epsilon_1+1)+2)+N)(\frac{1}{2} (\epsilon_1+1) (\alpha -\beta)+N) \]

\[ \Phi_{3}= 3\cdot 2^{39} \omega^2  (p+1-N) (\epsilon_1 (\alpha+\beta)+N+p+1)\]

\[ (\frac{1}{2} (\epsilon_1+1) (\alpha +\beta)+N)(\frac{1}{2} (\alpha  (\epsilon_1-1)+\beta \epsilon_1+\beta +2 N+2))\]

\[ \frac{1}{2} (\alpha (\epsilon -1)+\beta \epsilon_1+\beta +2 N-2)\frac{1}{2} (\epsilon_1 (\alpha+\beta)-\alpha+\beta +2 N) \]

\[ \frac{1}{2} (\epsilon_1 (\alpha+ \beta )+ \alpha -\beta +2 N+2)\frac{1}{2} (\epsilon  (\alpha +\beta )+\alpha-\beta +2 N) \]

\[ (\frac{1}{2} (\epsilon_1-1) (\alpha +\beta )+N)(\frac{1}{2} (\epsilon  (\alpha +\beta )+\alpha -\beta +2 N-2))\]
\end{document}